\documentclass[aps,prd,nofootinbib,longbibliography]{revtex4-1}
\usepackage{latexsym}
\usepackage{amsbsy}
\usepackage{mathrsfs}
\usepackage{color}

\usepackage{enumerate}

\usepackage{amsmath,amssymb,calc,amsfonts}

\usepackage{graphicx,calc,epsfig}
\usepackage[T1]{fontenc}
\usepackage[utf8]{inputenc}
\usepackage[english]{babel}
\usepackage{url}
\usepackage{amsthm}
\usepackage{hyperref}
\hypersetup{
    colorlinks=true, 
    linktoc=page,
    linkcolor=blue,  
    urlcolor=magenta
}
\usepackage{bbold}
\usepackage{bm}
\usepackage{booktabs}
\usepackage{braket}
\usepackage{caption}
\usepackage{enumerate}
\usepackage{enumitem}
\usepackage{fnpos} %
\usepackage[bottom,multiple]{footmisc} %controls position of footnotes
\usepackage{lipsum}	%come sopra %genera testo fittizio
\usepackage{mathtools} % oggetti matematici avanzati (abs, norma)
\usepackage{quoting}
\quotingsetup{font=small}
\usepackage{sidecap}
\usepackage{siunitx}
\usepackage{subfig}
\usepackage{textcomp}
\usepackage{tikz}
\usetikzlibrary{decorations.pathmorphing,calc}
\usepackage{titlesec}
\usepackage{tensor}
\usepackage{xfrac}

\newcommand{\be}{\begin{equation}}
\newcommand{\ee}{\end{equation}}
\newcommand{\bn}{\begin{equation*}}
\newcommand{\en}{\end{equation*}}
\newcommand{\ba}{\nopagebreak[3]\begin{eqnarray}}
\newcommand{\ea}{\end{eqnarray}}

\newcommand{\baa}{\nopagebreak[3]\begin{eqnarray*}}
\newcommand{\eaa}{\end{eqnarray*}}

\newcommand{\la}{\label}
\newcommand{\n}{\nonumber}

\newcommand{\va}{\scriptscriptstyle}

\newcommand{\lie}{\pounds} %Lie derivative symbol
\newcommand{\SU}{\mathrm{SU}}

\begin{document}

%\rmfamily

\title{Orthogonal gauge fixing of first order gravity}
\date{\today}
\author{Emanuele Alesci$^1$}
\email{eza69@psu.edu}
\author{Costantino Pacilio$^{2,3}$}
\email{cpacilio@sissa.it}
\author{Daniele Pranzetti$^4$}
\email{dpranzetti@perimeterinstitute.ca}
\affiliation{$^1$Institute for Gravitation and the Cosmos, Penn State, University Park, PA 16802, U.S.A.}
\affiliation{$^2$SISSA  - International School for Advanced Studies,  Trieste, Italy}
\affiliation{$^3$INFN, Sezione di Trieste, Trieste, Italy}
\affiliation{$^4$Perimeter Institute for Theoretical Physics,  Waterloo, ON N2L 2Y5, Canada}
\begin{abstract}
We consider  the first order connection formulation of 4D general relativity in the ``orthogonal'' gauge.
We show how the partial gauge fixing of the phase space canonical coordinates leads to the appearance of second class constraints in the theory. We employ the gauge unfixing procedure in order to successfully complete the Dirac treatment of the system.
While equivalent to the inversion of the Dirac matrix, the gauge unfixing allows us to work directly with the reduced phase space and the ordinary Poisson bracket. At the same time, we explicitly derive the new set of residual first class constraints preserving the partial gauge fixing, which are linear combinations of the original constraints, and these turn out to contain nonlinear terms. While providing an explicit example of how to consistently recast general relativity in a given partial gauge, the main motivation of this classical analysis is the application of the Quantum Reduced Loop Gravity program to a Schwarzschild black hole geometry.
\end{abstract}
\maketitle
%
%%%%%%%%%%%%%%%%%%%%%%%%%%%%%%%%%%%
\section{Introduction}\la{sec:Intro}

Einstein's theory of general relativity (GR) in its canonical formulation is a constrained system. The phase space is parametrized by the symmetric 3-metric tensor and its conjugate momentum components, which together amount to 6+6 local degrees of freedom. However, only 4 of them are physical. The extra 8 components represent gauge degrees of freedom and they are accounted by the 4 diffeomorphism constraints, 1 timelike and 3 spacelike, resulting from the canonical analysis of the Einstein-Hilbert action after the ADM decomposition of the spacetime manifold. The fact that each constraint kills 2 degrees of freedom follows from the first class nature of their algebra and it is related to the fact that the constraints are, at the same time, the generator of the local gauge symmetry (diffeo invariance in this case).

In the first order Ashtekar connection formulation \cite{Ashtekar:1991hf}, an extra gauge redundancy is introduced. In fact, in this case, the phase space configuration variable becomes a gauge connection and its conjugate momentum is a densitized triad (out of which the induced metric of the 3D spacelike hypersurface of the foliation is reconstructed) for a total of 18 degrees of freedom. The additional (with respect to the metric formulation) 6 components are taken care of by 3 extra first class constraints associated to the local rotational invariance of the triad, yielding again a total of 4 physical degrees of freedom. 

The presence of all this gauge symmetry is what makes it so difficult to find explicit general solutions of GR and the reason why physical applications are often times limited to symmetry reduced cases, where exact solutions representing good approximations to real physical situations can explicitly be found. In the quantum theory, these difficulties are further amplified by the presence of ordering ambiguities in the quantization procedure and anomalies in the resulting constraint algebra. Therefore, also in the quantum theory one would like to implement a symmetry reduction scheme for physical applications. 
For instance, quantum dynamics in canonical Loop Quantum Gravity (LQG) is implemented by the imposition of the Hamiltonian constraint on the states of the kinematical Hilbert space. However the Hamiltonian constraint is notoriously not tractable in practice, and this fact has stimulated investigations of symmetry reduced sectors of the theory, in which the dynamics may become more tractable.

However, the symmetry reduction strategy is crucially affected by an important choice: the order in which we perform the symmetry reduction and the quantization procedures. It is indeed well known that the two steps in general do {\it not} commute and the relation between the quantum theories outcome of the two alternative choices (first reduction and then quantization or the other way around) is often hard to assess. 

The easiest path to follow is usually the one of a classical symmetry reduction, since it is conceptually clearer (the notion of classical, continuum symmetry becomes often times fuzzy, or at least subtle, at the quantum level, where discrete structures emerge) and  it makes the quantization process considerably easier. This is indeed the choice commonly explored in canonical quantum gravity when one applies LQG techniques to cosmology and black hole physics (see \cite{Bojowald:2008zzb, Ashtekar:2011ni, FernandoBarbero:2010qy, DiazPolo:2011np, Frodden:2012en} and references therein). 
However, performing the symmetry reduction at the classical level usually hides the field theory aspects and it yields a quantum system with less degrees of freedom than the one obtained by the second choice of symmetry reduction at the quantum level following a Dirac approach (when this is possible to accomplish). This can cast some shadows on the reliability  of the physical results obtained through the former choice. In the context of LQG, these shadows manifest themselves in the ambiguity about the precise relation between Loop Quantum Cosmology and the full theory, in the cosmological case (see \cite{Bodendorfer:2014vea, Bodendorfer:2015hwl, Alesci:2014rra, Alesci:2015nja, Beetle:2016brg} for an ongoing investigation of this issue), and about the role of the Immirzi parameter in the recovery of the Bekenstein--Hawking entropy-area law, in the black hole case (for different proposals see, e.g., \cite{Jacobson:2007uj, Perez:2010pq, Engle:2011vf,  Frodden:2012dq, Ghosh:2013iwa,  Bodendorfer:2013hla, Pranzetti:2013lma, Ghosh:2014rra, Pranzetti:2014tla, Achour:2014eqa}). 
%Moreover, the fact that the gauge fixing in general does not commute with the rest of the constraints, making the whole system of second class, requires some extra care in  the  quantization process of a classically symmetry reduced system. 

Recently, motivated by its application in a LQG framework, a new strategy has been proposed which interpolates between the two alternatives of reduction first or quantization first.\footnote{Let us point out that an example of symmetry reduction implementation at the quantum level is provided by the Group Field Theory (GFT) reformulation of LQG in a second quantization language. This has been obtained   by modeling cosmological \cite{Gielen:2013kla, Gielen:2014uga, Calcagni:2014tga, Oriti:2016qtz} and black hole \cite{Oriti:2015qva, Oriti:2015rwa, Oriti:2018qty} quantum spacetimes in terms of quantum gravity condensates within the full theory. The results achieved through the construction of GFT condensates have  allowed to recover the outcome of the  previous LQG treatment and, at the same time, to clarify some of the ambiguities present in the  literature.} This new approach, dubbed `Quantum Reduced Loop Gravity' (QRLG), comprises two main steps, corresponding to a classical and a quantum analysis, and it was originally applied in a cosmological setting \cite{Alesci:2012md, Alesci:2013xd, Alesci:2014uha, Alesci:2014rra, Alesci:2015nja, Bilski:2015dra, Alesci:2016xqa, Alesci:2016gub, Bilski:2016pib}. 
We want to extend the program to the spherically symmetric  sector of GR in the first order connection formulation,  in order to apply it to the quantization of a Schwarzschild black hole geometry with LQG techniques. In this manuscript  we
concentrate on the first part of the analysis and we recast the classical phase space in a ``orthogonal'' gauge (defined below), compatible with a spherical symmetry reduction, by completing the Dirac analysis;
the quantization part will be carried out in a following work.

Before entering the technical part of the paper, let us summarize in a bit more detail the main aspects of our program.
Firstly, the partial gauge fixing conditions that one would like to impose in order to eventually implement the classical symmetry reduction are written down explicitly and added to the original set of constraints (this is done in Sec. \ref{sec:con}); 
we then study the algebra of this new set and, if second class constraints appear, we use the gauge unfixing procedure \cite{Mitra:1989fg, Mitra:1990mp, Anishetty:1992yk, Neto:2009rm} to impose them. This allows us to work directly with the reduced phase space\footnote{It is important to clarify that, at this stage, by ``reduced'' we intend ``partially gauge fixed'', {\it not} ``symmetry reduced''.}  and the ordinary Poisson bracket, however we will now be left with a new set of residual first class constraints preserving the gauge fixing, which will be linear combinations of the original constraints. As it will be explicitly shown in Sec.  \ref{sec:SCC}, the gauge unfixing procedure is completely equivalent to the Dirac bracket treatment of second class constraints; however, the price to pay is that the new form of the (first class) constraints left to impose will now be more complicated and, in particular, it will contain nonlocal terms (the explicit expressions are derived in Sec.  \ref{sec:extcon}). This is the most relevant result of the classical analysis performed here. 

In fact, the second part of our program does not intend to quantize the symmetry reduced phase space, but we will  start with the standard LQG quantization of the full GR phase space and proceed to the weak imposition of the gauge fixing conditions at the quantum level by means of the standard holonomy-flux algebra representation \cite{Ashtekar:1994mh, Ashtekar:1994wa, Ashtekar:1995zh}. 
This will yield the partially (orthogonal) gauge fixed kinematical Hilbert space of LQG.
The dynamics of the theory will now be encoded in the new expression of the Hamiltonian constraint obtained in the first part of the analysis; this will contain the gauge fixed version of the original Hamiltonian constraint plus extra, nonlinear terms which are fundamental in order to guarantee the consistency the of partial gauge fixing  procedure  under time evolution of the system. The main goal of this second part of the program is to obtain quantum corrections to physical semiclassical results by solving the evolution equations for initial data. At a first level of approximation, such effective equations can be obtained through expectation values on coherent states constructed out of reduced spin network states adapted to our choice of gauge fixing and encoding the information of a given semiclassical geometry.
It is thus at the level of the quantum states that the symmetry reduction is implemented.\footnote{The first complete treatment of a vacuum Schwarzschild spacetime in a geometrodynamical setting is due to the seminal work of Kucha\v{r} \cite{Kuchar:1994zk}, although a previous canonical analysis in the Ashtekar formalism was performed by Thiemann and Kastrup in \cite{Kastrup:1993br}, containing very similar results for the parametrization of the {\it symmetry} (see footnote 2) reduced phase space.}

The gauge unfixing procedure for the ``radial''  gauge applied to the case of spherical symmetry,  as well as full 4d general relativity  in the metric formulation was previously considered in \cite{Bodendorfer:2014wea, Bodendorfer:2015aca}. While our implementation of the gauge unfixing procedure closely parallels that of \cite{Bodendorfer:2015aca}, the main difference is represented by our use of Ashtekhar variables and in how we gauge fix the radial sector of the spatial metric. 
In fact, what the authors call `radial' gauge in these works is slightly, but crucially (for the resulting final form of the Hamiltonian constraint) different from our gauge choice. That is why, to avoid confusion with previous literature, we have decided to refer to our gauge choice as ``orthogonal'' rather than ``radial''. The construction of connection variables for the spherically symmetric case was sketched in \cite{Bodendorfer:2014wea} and then investigated in much more detail in \cite{Bodendorfer:2015qie} (see also Appendix B of \cite{Bodendorfer:2015hwl}), in order to apply  LQG techniques  to implement reduction to spherical symmetry at the quantum level. In these other works the authors
introduce a Peldan hybrid spin connection, different from the
Ashtekar--Barbero connection considered here, and they partially relax the gauge restriction on the radial part of the metric with respect to the one of  \cite{Bodendorfer:2014wea, Bodendorfer:2015aca}. This choice of radial gauge is more similar to the one adopted in this work. However, the authors of  \cite{Bodendorfer:2015qie} as well as  \cite{Bodendorfer:2015hwl} introduce an extra gauge condition on the radial shift, imposing that this does not depend on the angular coordinates; in this way, the correction terms to the radial diffeo constraint are not computed explicitly since those would depend exactly on the angular derivatives of the radial shift, as  will be the case also in our analysis (see the results of Sec.  \ref{sec:extcon}). Moreover, the Hamiltonian constraint is not included in the analysis of those works and no extended version for it is derived.
Therefore, it is not possible for us to explicitly compare our results with those of  \cite{Bodendorfer:2015qie, Bodendorfer:2015hwl}. 
In light of these differences with previous applications of the gauge unfixing procedure to implement some version of partial gauge  in full GR, the results we derive in Sec.  \ref{sec:extcon} for the extended radial diffeo and Hamiltonian constraints represent the main original results of this manuscript.
%which yields some important differences in the quantum theory with respect to the 

Concerning the quantization scheme of   \cite{Bodendorfer:2015qie} for a spherically symmetric spacetime, the use of a Peldan
hybrid spin connection leads to the construction of a kinematical Hilbert space where techniques of the full LQG framework are applied, but still relying also on the notion of point holonomy for some of the degrees of freedom. Point holonomies are used also in  \cite{Bodendorfer:2015hwl} in order to quantize some of the phase space configuration variables (although different ones with respect to \cite{Bodendorfer:2015qie}). This allows for some technical simplifications in the quantum theory, yielding for instance a diagonal volume operator, but it represents as well a departure from the standard LQG Hilbert space built on a full $\SU(2)$ Ashtekar-Barbero connection. This different kinematical structure would eventually reflect on the kind of quantum corrections that can be derived for an effective Hamiltonian, for instance.
We are not going to present any result concerning the application of our classical analysis of the gauge unfixing procedure performed here to the quantum reduction of full LQG to spherical symmetry (see Sec.  \ref{sec:Conc} though, for some details on our quantization strategy and \cite{Alesci:2018loi} for its explicit implementation); however, we anticipate that application of QRLG techniques to a spherically symmetric spatial manifold will still rely on the $\SU(2)$ Ashtekar-Barbero connection for all of the kinematical degrees of freedom, with proper restrictions  applied in order to implement the quantum reduction. In this way, we still have only $\SU(2)$ holonomies, although just a restricted set of representation matrix elements will be allowed, so that the  reduced flux operators become diagonal in the QRLG Hilbert space for our orthogonal gauge. 
Similarly to the formulation of  \cite{Bodendorfer:2015qie},
this has the advantage of greatly simplifying calculations involving the Hamiltonian constraint operator. However, since in the QRLG construction we will not have to rely on point holonomies, there will be more degrees of freedom captured by the reduced kinematical Hilbert space, making our construction closer to the one of the full theory and yielding different quantum corrections in the effective dynamics.
The classical investigation performed here is tailored for this  briefly sketched 
 quantum construction, which differs from previous attempts; this thus provides further  motivation for the analysis of this manuscript. 
We will spell out and comment on these differences more in detail at several points through the paper.

Let us stress  that, while our main motivation is to  apply the results obtained here to the LQG quantization of a black hole \cite{Alesci:2018loi}, the classical analysis we perform is interesting on its own, since it represents a successful treatment of a second class Hamiltonian system according to the Dirac procedure, allowing us to recast  full 4D general relativity in the first order formulation in a partial gauge. 
%---The gauge unfixing procedure for the radial gauge applied to the case of spherical symmetry reduction was previously considered in the metric formulation of GR in
% \cite{Bodendorfer:2014wea, Bodendorfer:2015aca, Bodendorfer:2015qie}.

%%%%%%%%%%%%%%%%%%%%%%%%%%%%%%%%%%%
\section{Constraints and gauge conditions}\la{sec:con}
We want to impose gauge conditions in vacuum GR compatible with a reduction to spherical symmetry. 
Let us assume that the spacetime admits a foliation by smooth 3D hypersurfaces $\Sigma_t$.
We will work in the Ashtekar canonical formulation of vacuum GR, in which, after imposition of the time gauge, the action takes the form
\be
\label{eq:action:1}
S=\frac{1}{16\pi G}\int dt\int_{\Sigma_t} d^3x \left[\frac{2}{\gamma}E^a_i\lie_tA_a^i-NH-N^aV_a-\Lambda^iG_i\right],
\ee
where $\gamma$ is the Immirzi parameter.
The action \eqref{eq:action:1} defines the phase space coordinates in terms of an $\SU(2)$ connection configuration variable $A$ and its conjugate momentum $E$ (densitized triad), and it describes a pure constraint theory, with $N, N^a, \Lambda^i$ playing the role of Lagrange multipliers. The explicit expressions of the constraints are
\begin{widetext}
\begin{subequations}
\label{eq:constraints:1}
\begin{align}
&G_i=\partial_a E^a_i+\epsilon\indices{_{ij}^k}A_a^jE_k^a\,, &\text{Gauss constraint}\\
&V_a=F_{ab}^iE_i^b\,, &\text{Vector constraint}\\
&H=\frac{\gamma E_i^aE_j^b}{2\sqrt{\text{det}(E)}}\left[\epsilon\indices{^i^j_k}F_{ab}^k-2(1+\gamma^2)K^i_{[a}K^j_{b]}\right] \,,&\text{Hamiltonian constraint}
\end{align}
\end{subequations}
\end{widetext}
where
\be
\label{eq:f:1}
F_{ab}^i=\partial_aA_b^i-\partial_bA_a^i+\epsilon\indices{^i_{jk}}A_a^jA_b^k
\ee
is the curvature of the Ashtekar connection $A_a^i$.

Let us now introduce a local set of coordinates to parametrize a neighborhood of a point in a given constant time slice $\Sigma_t$.
Relying on the geometrical construction of \cite{Duch:2014hfa},
%The properties of the time slices $\Sigma_t$ in \eqref{eq:action:1} are so far generic. We assume that $\Sigma_t$ is foliated as $\Sigma_t\sim\mathbb{R}\times\mathcal{B}$. The topology of $\mathcal{B}$ is irrelevant for most of the following treatment, but for definiteness we specify it to the case we are interested in, namely $\mathcal{B}\sim S^2$. 
 we coordinate $\Sigma_t$ by spherical coordinates $(r,\theta,\phi)$. Such a  set of coordinates, which relies of the use of radial geodesics, can always be defined locally and, in general, they can only take value in a finite range. In the following, we do not need to specify the finite interval for the angular coordinates and we assume the radial coordinate to take values in the finite range $r\in[0, \bar r]$ (in the case of a spherically symmetric geometry one can extend the validity of these spherical coordinates to their full range, up to nontrivial topologies).
 Moreover we make the further restricting requirement that the radial evolution vector has vanishing shift; this implies that $r^a$, the unit spacelike radial vector, is proportional to $\delta^{ar}$.

Given the above setup, the spatial index $a$ takes values $a=r,\theta,\phi$, and the integration element in \eqref{eq:action:1} is $d^3x=dr\,d\theta\,d\phi$. The $\SU(2)$ internal index $i$ takes, as usual, values $i=1,2,3$. The canonical Poisson brackets (PB) induced by \eqref{eq:action:1} are
\be
\label{eq:poisson:1}
\{A_a^i(\vec{x}),E^b_j(\vec{y})\}=8\pi G\gamma\delta^b_a\delta^i_j\delta(\vec{x}-\vec{y})\,,
\ee
where $\delta(\vec{x}-\vec{y})=\delta(r_x-r_y)\delta(\theta_x-\theta_y)\delta(\phi_x-\phi_y)$. The algebra of the constraints determined by \eqref{eq:poisson:1} turns out to be first class.

We want to fix the system in a gauge conveniently adapted to the foliation of $\Sigma_t$. We choose an ``orthogonal'' gauge defined by $E_3^a$ being aligned with $r^a$, which, by the previous discussion, is equivalent to require
\begin{subequations}
\label{eq:gauge:1}
\begin{align}
&E_I^r=0\,,\quad I=1,2\,,\\
&E_3^A=0\,,\quad A=\theta,\phi\,,
\end{align}
\end{subequations}
where we made a decomposition along radial and tangential indices. In particular, we use capital letters $I,J,K,\dots$ to label internal indices $1, 2$. Similarly, we use capital letters $A,B,C,\dots$ to label tangential coordinates $\theta, \phi$. We can understand Eqs. \eqref{eq:gauge:1} as a set of four gauge conditions for our original theory \eqref{eq:action:1}.

The block-diagonal structure of the gauge choice \eqref{eq:gauge:1} can be better appreciated by rewriting the fluxes in a matricial form with internal indices $3,I$ labelling rows and space indices labelling columns, namely
\be
\begin{bmatrix}
E_3^r & 0 & 0 \\
0 & E_1^\theta & E_1^\phi \\
0 & E_2^\theta & E_2^\phi 
\end{bmatrix}
\ee 

It is then evident the similarity with the radial gauge choice structure of the  spatial metric $h_{ab}$ adopted in \cite{Bodendorfer:2014wea, Bodendorfer:2015aca}, where $h_{ab}$ is a block diagonal 3x3 matrix of the form
\be
\begin{bmatrix}
h_{rr} & 0 & 0 \\
0 & h_{\theta\theta} & h_{\theta\phi}  \\
0 & h_{\phi\theta} & h_{\phi\phi}
\end{bmatrix}\,.
\ee

However, the block-diagonal structure \eqref{eq:gauge:1} leaves more freedom than the conventional ``radial'' gauge considered in \cite{Bodendorfer:2014wea, Bodendorfer:2015aca}, in which the component $h_{rr}$ is fixed to 1. In fact, $E^r_3$ is left unconstrained and thus $h_{rr}$ is still a degree of freedom in our constraint system\footnote{The condition $h_{rr}=1$ implies, in terms of fluxes, $E^r_3=\epsilon_{3}\!^{IJ}E^\theta_I E^\phi_J$.}. As we will point out below, this apparently minor difference in gauge choice can actually lead to quite different extended Hamiltonian constraint in the GU procedure, thus a comparison with the analysis of \cite{Bodendorfer:2014wea, Bodendorfer:2015aca} in the general full GR case is not straightforward (however, we will comment on the differences when specializing to the spherically symmetric case at the end of the paper).
In light of these differences, we use the expression ``orthogonal'' gauge to denote the block-diagonal in the sense explained above.

We must check the PB algebra between the gauge conditions \eqref{eq:gauge:1} and the constraints \eqref{eq:constraints:1}. To this aim, it is convenient to replace the vector constraint $V_a$ with the diffeomorphisms constraint
\be
\label{eq:constraints:2}
H_a=V_a-A^i_aG_i\,,
\ee
which generates spatial diffeomorphisms on $\Sigma_t$:\footnote{From now on we work in units $8\pi G=1$.}
\begin{subequations}
\begin{align}
&\{E_i^a,\vec{H}[\vec{N}]\}=\gamma\lie_{\vec{N}}E_i^a=\gamma\left(N^b\partial_bE_i^a-E^b_i\partial_bN^a+\partial_bN^b\,E_i^a\right) \label{eq:poisson:2}\,,\\
&\{A_a^i,\vec{H}[\vec{N}]\}=\gamma\lie_{\vec{N}}A^i_a=\gamma\left(N^b\partial_bA^i_a+A_b^i\partial_aN^b\right)\,,
\end{align}
\end{subequations}
where $A_a^i$ transforms as an ordinary covector while $E^a_i$ transforms as a vector density. Here $\vec{H}[\vec{N}]$ denotes the smeared diffeomorphisms constraint
\be
\label{eq:constraints:3}
\vec{H}[\vec{N}]=\int d^3x N^a\,H_a.
\ee
In order to facilitate the computation of the PB between the constraints and the gauge conditions, we adopt the following notation: We denote by $\vec{N^a}$ a smearing vector field having a nonvanishing component only along the $a$-th direction, $(N^a)^b=\eta^a\delta^{ab}$.\footnote{As usual, indices in the same positions are not summed over, unless otherwise specified.} Correspondingly the smearing $\vec{H}[\vec{N^a}]$ selects only the $a$-th component of $H_b$; for example, $N^\theta\equiv(0,\eta^\theta,0)$, and $\vec{H}[\vec{N^\theta}]\equiv\int \eta^\theta H_\theta$.
Similarly we denote by $\vec{\Lambda^i}$ a vector in the internal space with the nonvanishing component only along the $i$-th internal direction, $(\Lambda^i)^j=\lambda^i\delta^{ij}$. Therefore, $\vec{G}[\vec{\Lambda^i}]$ selects only the $i$-th component of $G_j$; for example, $\Lambda^1\equiv(\lambda^1,0,0)$ and $\vec{G}[\vec{\Lambda^1}]\equiv\int\lambda^1G_1$.

With these conventions we find that, on the gauge surface selected by \eqref{eq:gauge:1}, \eqref{eq:poisson:2} gives\footnote{In this paper we assume vanishing boundary conditions for the smearing functions.}
\begin{subequations}
\label{eq:poisson:3}
\begin{align}
&\{E_I^r,\vec{H}[\vec{N^A}]\}\approx-\gamma E_I^B\partial_B\eta^A\delta^{Ar}=0\,,\\
&\{E_I^r,\vec{H}[\vec{N^r}]\}\approx-\gamma E_I^A\partial_A\eta^r \label{eq:poisson:3:2}\,,
\end{align}
\end{subequations}
and
\begin{subequations}
\label{eq:poisson:4}
\begin{align}
&\{E_3^A,\vec{H}[\vec{N^B}]\}\approx-\gamma E_3^r\partial_r\eta^B\delta^{AB}\,,\\
&\{E_3^A,\vec{H}[\vec{N^r}]\}\approx-\gamma E_3^r\partial_r\eta^r\delta^{Ar}=0\,,
\end{align}
\end{subequations}
where the symbol $\approx$ denotes projection of the phase space onto the gauge surface \eqref{eq:gauge:1}. We thus see that $E_I^r$ is second class only with $H_r$, while $E_3^A$ is second class only with $H_A$.

Regarding the Gauss constraint, $G_3$ is first class with both $E_I^r$ and $E_3^A$. This was already expected from the geometrical meaning of $G_3$, since it generates internal rotations orthogonal to the third internal direction.
On the other hand, we have
\begin{subequations}
\label{eq:poisson:5}
\begin{align}
&\{E_I^r,\vec{G}[\vec{\Lambda^J}]\}\approx-\gamma\lambda^J\epsilon\indices{^J_I}E_3^r\,,\\
&\{E_3^A,\vec{G}[\vec{\Lambda^J}]\}\approx\gamma\lambda^J\epsilon\indices{^J^I}E_I^A\,,
\end{align}
\end{subequations}
meaning that both $E_I^r$ and $E_3^A$ are second class with $G_I$.

The PB between the gauge conditions and the Hamiltonian constraint are not explicitly needed in the rest of the paper, but we 
show them here just for completeness. They read
\begin{subequations}
\begin{align}
&\{E_I^r,H[N]\}\approx -\gamma^2 \epsilon^{J}\!_I \partial_A\left( \frac{N E^A_J E^r_3}{\sqrt{\text{det}(E)}}\right)
+\gamma^2  \frac{N E^A_I E^r_3}{\sqrt{\text{det}(E)}}\left(A^3_A-\frac{(1+\gamma^2)}{\gamma}K^3_A\right)\,,\\
&\{E_3^A,H[N]\}\approx-\gamma^2 \epsilon^{IJ} \partial_B\left( \frac{N E^A_I E^B_J}{\sqrt{\text{det}(E)}}\right)
+\gamma^2  \frac{N E^A_I E^r_3}{\sqrt{\text{det}(E)}}\left(A^I_r-\frac{(1+\gamma^2)}{\gamma}K^I_r\right)\,.
\end{align}
\end{subequations}

%%%%%%%%%%%%%%%%%%%%%%%%%%%%%%%%%%%
\section{Implementation of the second class constraints} \la{sec:SCC}
%%%%%%%%%%%%%%%
The treatment of a second class Hamiltonian system follows the Dirac procedure \cite{dirac1964lectures}. This consists of splitting the set of the original constraints \emph{and} of the gauge conditions, all of which we collectively refer to as ``the constraints'', in two subsets: the first class subset, consisting of those constraints that commute with each other and with the second class constraints; and the second class subset, in which every member does not commute with at least another one.

There is some ambiguity in this splitting. However, it is clear that, in order to preserve the number of physical degrees of freedom of the phase space, the second class constraints must be twice as many as the gauge conditions. In our case this implies that, since \eqref{eq:gauge:1} are four conditions, four and only four out of the original seven constraints $G_i$, $H_a$ and $H$ are second class with them.

In turn, this leaves three residual first class constraints. They do not necessarily coincide \emph{directly} with three constraints from the initial set, but they can come in linear combinations with the others (this is the source of the splitting ambiguity). Indeed, from \eqref{eq:poisson:3}-\eqref{eq:poisson:5}, only $G_3$ is directly first class. Therefore, the remaining two first class constraints must be expressed as linear combinations of the original ones.

Once this splitting is completed, one must invert the Dirac matrix, i.e. the antisymmetric matrix whose elements are the PB of the second class constraints. The inverse of the Dirac matrix then allows us to implement the second class constraints by deforming the Poisson brackets into the so called Dirac brackets. The remaining first class constraints and the dynamics of the theory can be finally imposed with the Dirac brackets.

However, finding a representation of the Dirac brackets can be problematic, introducing serious obstructions to the completion of the  quantization process. It is hence useful to follow an alternative, but equivalent, route to impose the second class constraints. One possibility is represented by the so-called ``gauge unfixing'' (GU) procedure  introduced in \cite{Mitra:1989fg, Mitra:1990mp, Anishetty:1992yk}  (see also \cite{Neto:2009rm}). The advantage of the GU is that one works directly with the reduced phase space variables, while still using the ordinary Poisson brackets. Moreover, it gives a direct way to compute the gauge invariant residual first class constraints.

\subsection{The gauge unfixing procedure}
The GU consists of finding an extension of the  phase space invariant under the flow of the gauge conditions. In the case of \eqref{eq:gauge:1}, this amounts to finding extensions of $A_r^I$ and $A_A^3$. To avoid confusion, these extensions are denoted with a tilde: $\tilde{A}_r^I$ and $\tilde{A}_A^3$. They are obtained by adding to $A_r^I$ and $A_A^3$ terms proportional to the original constraints.

Before going into the details, let us explain the procedure in a more formal way.  
Our application of the GU procedure is somehow the reverse of what is usually done. Usually, the GU is applied to an original second class system of constraints in order to turn a subset of them into a first class system. In our case, we start with a first class system and we transform it into an auxiliary second class one by imposing a set of 
gauge fixing conditions for some of the phase space coordinates. At this point, by applying the GU procedure to the auxiliary second class system we can obtain a new first class system, in which we have traded some of the original constraints with the gauge fixing conditions that we have chosen.

More precisely, let $Q_a$ and $P^a$ be, respectively, the configuration and momentum fields of our field theory with Poisson brackets
\be
\label{eq:poisson:6}
\{P^a(\vec{x}),Q_b(\vec{y})\}=\gamma\,\delta^a_b\delta(\vec{x}-\vec{y})\,,
\ee
where now $a,b,c,\dots$ stand both for internal and tangential indices.

The theory is supposed to be equipped with a set of first class constraints $\{V_i\}$:
\be
\{V_i(\vec{x}),V_j(\vec{y})\}=0\,,
\ee
where $i,j,k,\dots$ are constraint labels.

We impose as gauge conditions
\be
\label{eq:gauge:2}
\chi_a\approx 0\,,
\ee
 that a subset  of the configuration fields $\{Q_a\}$ vanishes.  
 The enlarged set of constraints $\{V_i , \chi_a \}$ is now
second class. At this point we run the GU machinery to turn the $\{\chi_a \}$ into first class constraints, while interpreting  a subset $\{C_i\}$ of equal number of the original constraints $\{V_i\}$ as gauge conditions for the $\{\chi_a \}$.
 
 In order to do so, we  have to find gauge invariant extensions of the corresponding momenta $\{P_\chi^a\}$. Let $\tilde{P}_\chi^a$ be
\be
\label{eq:p:tilde}
\tilde{P}_\chi^a(\vec{x})=P_\chi^a(\vec{x})+\int d\vec{y}\,C_i(\vec{y})\mathbb{N}^{ia}(\vec{y},\vec{x})+\dots\,,
\ee
where  the dots indicate terms of higher powers of the $C_i$'s.
In \eqref{eq:p:tilde}
$\mathbb{N}^{ia}$ is a distributional matrix and, together with its higher power counterparts, it must be fixed by requiring the gauge invariance of $\tilde{P}_\chi^a$, i.e. 
\be
\label{eq:poisson:7}
\{\chi_a(\vec{x}),\tilde{P}_\chi^b(\vec{y})\}\approx0.
\ee
Finally, by replacing $P_\chi^a$ with $\tilde P_\chi^a$ in the other remaining constraints, we manage to promote the auxiliary second class constraints $\{V_i , \chi_a \}$ to a new first class set.

In general, imposing \eqref{eq:poisson:7} gives recursive relations for $\mathbb{N}^{ia}$, and for its higher power counterparts, that are not easy to solve. However a great simplification occurs when the $C_i$'s depend on the momenta $P_\chi$ at most linearly: in this case the  higher power terms in \eqref{eq:p:tilde} drop out, and $\mathbb{N}^{ia}$ becomes independent of the $P_\chi$'s. We will see in a moment a direct example of such simplifications.
In fact, observe that the constraints $G_i$ and $H_a$ are all linearly dependent on the fields $A_a^i$; therefore, if we choose the $C_i$'s among them, as we will actually do, these simplifications apply. This is the main reason why we replaced the vector constraint $V_a$ with the diffeomorphisms constraint $H_a$.

\subsection{Extended phase space}

With these simplifications in mind, combining \eqref{eq:p:tilde} and \eqref{eq:poisson:7} we obtain
\be
\label{eq:tilde:2}
0\approx-\gamma\,\delta^a_b\delta(\vec{x}-\vec{z})+\int d\vec{y}\,\{\chi_b(\vec{z}),C_i(\vec{y})\}\mathbb{N}^{ia}(\vec{y},\vec{x})\,,
\ee
 from which we see that $\mathbb{N}^{ia}$ is the inverse of the matrix
 \be
 \mathbb{A}_{ai}=\gamma^{-1}\{\chi_a(\vec{z}),C_i(\vec{y})\}.
 \ee
 The application of the GU procedure thus boils down to finding the inverse matrix $(\mathbb{A}^{-1})^{ia}$ and replacing $\mathbb{N}^{ia}=(\mathbb{A}^{-1})^{ia}$ inside \eqref{eq:p:tilde}.
 Finally, promoting $P_\chi^a$ to $\tilde{P}_{\chi}^a$, we end up with a theory invariant under the gauge conditions, and we can work only with the physical degrees of freedom and the eventual gauge residual ones.
 
 Notice that, once the replacement $P_\chi^a\to\tilde{P}_{\chi}^a$ is performed inside the remaining constraints, these are mapped into linear combinations of the original ones with the $C_i$'s. This is a direct way of obtaining the true gauge invariant first class constraints.

In order to invert the matrix $\mathbb{A}_{ai}$, it is convenient to define its smeared version
\be
\mathbb{A}(\vec{x},\alpha)_{ai}=\gamma^{-1}\{\chi_a(\vec{x}),C_i[\alpha]\}\,,
\ee
where all the $C_i$'s are smeared with the same smearing function $\alpha(\vec{x})$. Then the inverse $(\mathbb{A}^{-1})^{ia}$ is the matrix such that
\be
\int d\vec{y} (\mathbb{A}^{-1})^{ia}(\vec{x},\vec{y})\mathbb{A}(\vec{y},\alpha)_{aj}=\delta^i_j\alpha(\vec{x}).
\ee

Let us now have a closer look at the matrix $\mathbb{A}_{ai}$ and show how to invert it in the case of interest described in the previous section. First of all, we must choose the constraints $C_i$. Guided by the physical meaning of the constraints, we observe that, to implement the gauge, two rotations generated by $G_1$ and $G_2$ align $r^I$ along the third internal axis, while two diffeomorphisms generated by $H_\theta$ and $H_\phi$ make the angular components of $r^a$  vanish. Henceforth, we choose ${C_i}={G_I,H_A}$. The matrix $\mathbb{A}(\vec{x},\alpha)_{ai}$ then becomes
\be
\label{eq:dirac:1}
\mathbb{A}(\vec{x},\alpha)=
\begin{bmatrix}
c\indices{^A_J} & a\indices{^A_B}\\
b\indices{_I_J}& \emptyset\indices{_I_B}
\end{bmatrix}\,,
\ee
where
\begin{subequations}
\begin{align}
&a\indices{^A_B}(\vec{x})=\left\{E_3^A(\vec{x}), H_B[\alpha]\right\}=-E_3^r(\vec{x})\partial_r\alpha(\vec{x})\delta\indices{^A_B} \label{eq:small:a}\,,\\
&b\indices{_I_J}(\vec{x})=\left\{E_I^r(\vec{x}), G_J[\alpha]\right\}=\alpha(\vec{x})\epsilon_{IJ}E^r_3(\vec{x}) \label{eq:small:b}\,,\\
&c\indices{^A_J}(\vec{x})=\left\{E_3^A(\vec{x}),G_J[\alpha]\right\}=\alpha(\vec{x})\epsilon\indices{_J^K}E_K^A(\vec{x}) \label{eq:small:c}\,,\\
&\emptyset\indices{_I_B}(\vec{x})=\left\{E_I^r(\vec{x}), H_B[\alpha]\right\}=0\,,
\end{align}
\end{subequations}
and we used Eqs.\,\eqref{eq:poisson:3}-\eqref{eq:poisson:5}.

The inverse matrix $(\mathbb{A}^{-1})^{ia}$, that we derive in the Appendix \ref{sec:app:a}, reads
\be
\label{eq:dirac:4}
\mathbb{A}^{-1}(\vec{x},\vec{y})=
\begin{bmatrix}
\emptyset\indices{^I_B}&(b^{-1})\indices{^I^J}\\
(a^{-1})\indices{^A_B}&d\indices{^A^J}
\end{bmatrix}\,,
\ee
where
\begin{widetext}
\begin{subequations}
\begin{align}
&(a^{-1})\indices{^A_B}(\vec{x},\vec{y})=\frac{\delta\indices{^A_B}}{E_3^r(\vec{y})}\,\Theta(r_y-r_x)\,\delta(\theta_x-\theta_y)\,\delta(\phi_x-\phi_y) \label{eq:small:amo}\,,\\
&(b^{-1})\indices{^I^J}(\vec{x},\vec{y})=-\frac{\epsilon\indices{^I^J}}{E_3^r(\vec{y})}\,\delta(\vec{x}-\vec{y}) \label{eq:small:bmo}\,,\\
&d\indices{^A^J}(\vec{x},\vec{y})=\frac{\delta^{JK}E^A_K(\vec{y})}{\left(E_3^r(\vec{y})\right)^2}\,\Theta(r_y-r_x)\,\delta(\theta_x-\theta_y)\,\delta(\phi_x-\phi_y) \label{eq:dmo}\,,
\end{align}
\end{subequations}
\end{widetext}
and $\Theta$ is the Heaviside step distribution.

We can now compute the extended momenta:
\ba
\label{eq:a:tilde:1}
\tilde{A}_A^3(\vec{x})&=&A_A^3(\vec{x})+\int d\vec{y}\,H_B(\vec{y})(a^{-1})\indices{^B_A}(\vec{y},\vec{x})\n\\
&=&A_A^3(\vec{x})+\frac{1}{E_3^r(\vec{x})}\int dr'\,H_A(r')\Theta(r-r')\n\\
&=&\frac{1}{E^r_3(\vec{x})}\int_0^r dr'\left[D_A+E_3^r\partial_AA_r^3\right]_{r'}\,,
\ea
where we have defined
\be
D_A\equiv E^B_I\partial_AA_B^I-\partial_B\left(A_A^IE^B_I\right)\,.
\ee
In the last step of \eqref{eq:a:tilde:1} we have used the boundary condition $A_A^3(r=0,\theta,\phi)=0$.

Similarly,
\ba
\label{eq:a:tilde:2}
\tilde{A}_r^I(\vec{x})&=&A_r^I(\vec{x})+\int d\vec{y}\,H_A(\vec{y})d\indices{^A^I}(\vec{y},\vec{x})
+\int d\vec{y}\,G_J(\vec{y})\left(b^{-1}\right)^{JI}(\vec{y},\vec{x})\n\\
&=&A_r^I(\vec{x})+\frac{\delta^{IJ}E^A_J(\vec{x})}{\left(E_3^r(\vec{x})\right)^2}\int_0^r dr'\,H_A(r')
+\frac{\epsilon^{IJ}}{E_3^r(\vec{x})}G_J(\vec{x})\n\\
&=&\frac{\epsilon^{IJ}\partial_AE^A_J(\vec{x})}{E^r_3(\vec{x})}+\frac{\delta^{IJ}E^A_J(\vec{x})}{\left(E_3^r(\vec{x})\right)^2}\int_0^r dr'\left[D_A+E_3^r\partial_AA_r^3\right]_{r'}\,,
\ea
where again in the last step we have used $A_A^3(r=0,\theta,\phi)=0$.

We have thus obtained the extended phase space. The next step consists of replacing \eqref{eq:a:tilde:1} and \eqref{eq:a:tilde:2} into the remaining constraints, in order to generate their extended representation.

Before going on, let us observe that Eqs.\,\eqref{eq:a:tilde:1} and \eqref{eq:a:tilde:2} are equivalent to solve directly the constraints on the gauge surface, i.e.
\be
H_A\approx 0\implies A_A^3(\vec{x})\approx\frac{1}{E^r_3(\vec{x})}\int_0^r dr'\left[D_A+E_3^r\partial_AA_r^3\right]_{r'}\,,\\
\ee
which, in turn, implies
\be
G_I\approx 0 \implies A_r^I \approx\frac{\epsilon^{IJ}\partial_AE^A_J(\vec{x})}{E^r_3(\vec{x})}
 +\frac{\delta^{IJ}E^A_J(\vec{x})}{\left(E_3^r(\vec{x})\right)^2}\int_0^r dr'\left[D_A+E_3^r\partial_AA_r^3\right]_{r'}\,.
\ee
However, the main advantage of the GU with respect to the direct solution of the second class constraints is the possibility to obtain the expression of the gauge invariant operators in a straightforward manner. Indeed, through the replacement $P_\chi^a\to\tilde{P}\indices{_{\chi}^a}$ and using \eqref{eq:p:tilde}, it is easy to distinguish the original operator from the corrections induced by the requirement of gauge invariance, which are proportional to the second class constraints.

Moreover, notice also that we have the freedom to choose the Dirac matrix as
\be
\label{eq:dirac:2}
\mathbb{D}=
\begin{bmatrix}
\emptyset&\mathbb{A}\\
-\mathbb{A}^T&\emptyset
\end{bmatrix}\,,
\ee
where $\mathbb{A}$ is the same as in \eqref{eq:dirac:2}. Indeed, this corresponds to select $G_I$ and $H_A$ as the second class constraints.

We easily see that such a choice is compatible with the counting of the phase space degrees of freedom. In the ungauged original theory one starts with 18 phase space degrees of freedom minus 2$\times$(7 first class constraints), which gives 4 physical degrees of freedom.\footnote{Recall that a first class constraint freezes two phase space degrees of freedom, while a second class constraint freezes only one.} When we impose the gauge fixing, we have 18 degrees of freedom minus 2$\times$(3 irreducible first class constraints) minus 4 second class constraints minus 4 gauge conditions, which gives again 4 physical degrees of freedom.

Therefore the GU procedure is equivalent to the inversion of the Dirac matrix. In this case, the main advantage of the GU is that the Poisson brackets are not modified, while in the Dirac method the correction of the Poisson brackets makes it harder to implement them at the quantum level.
%%%%%%%%%%%%%%%%%%
\section{Extended representation of the remaining constraints}\la{sec:extcon}
The extended representation of the remaining constraints ($G_3$, $H_r$ and $H$) is obtained from the original ones, by promoting $A_A^3$ and $A_r^I$ to their extended versions $\tilde{A}_A^3$ and $\tilde{A}_r^I$, and specifying the result to the gauge surface \eqref{eq:gauge:1}.

The Gauss constraint $G_3[\Lambda^3]$ is not affected, as it is clear from its geometrical meaning.

The radial diffeomorphism constraint $H_r[N^r]$ acquires extra terms in the form of linear combinations of $H_A$ and $G_I$, namely
\be
\label{eq:ext:1}
\tilde{H}_r[N^r]\approx H_r[N^r]+H_A[\gamma^A] + G_I[\gamma^I]\,,
\ee
where
\ba
\label{eq:g:1}
\gamma^A&=&\int d\vec{r'}\left(E_I^B\partial_BN^r\right)_{r'}\,d^{AI}(\vec{r'},\vec{x})\n\\
&=&\int_r^{\bar{r}} dr'\left(\frac{\delta^{IJ}E_I^AE_J^B}{(E_3^r)^2}\partial_BN^r\right)_{r'}\,,
\ea
and
\ba
\label{eq:g:2}
\gamma^I&=&\int d\vec{r'}\left(E_J^B\partial_BN^r\right)_{r'}\,(b^{-1})^{IJ}(\vec{r'},\vec{x})\n\\
&=&-\frac{\epsilon^{IJ}E_J^A}{E_3^r}\partial_AN^r\,.
\ea
Alternatively, let us define the {\it reduced radial diffeomorphisms} $\mathcal{H}_r$, consisting of those parts of $H_r$ that do not contain $A_A^3$ and $A_r^I$, explicitly
\be
\label{eq:barh}
\mathcal{H}_r=\left(\partial_rA_A^I\right)E_I^A-A_r^3\partial_rE_3^r.
\ee
Then, using the last line of \eqref{eq:a:tilde:2}, we obtain
\be
\label{eq:hr:tilde}
\tilde{H}_r[N^r]\approx\mathcal{H}_r[N^r]+\int d\vec{x}\,\left(\partial_AN^r\right)\left[\frac{\epsilon^{IJ}E_I^A\partial_BE_J^B}{E_3^r} 
+\frac{\delta^{IJ}E_I^AE_J^B\mathcal{I}_B}{(E_3^r)^2}\right]
\ee
where, to shorten the notation, we have defined
\be
\mathcal{I}_A\equiv \int_0^r dr'\left[D_A+E_3^r\partial_A A_r^3\right]_{r'}.
\ee

Notice that the extension of the radial diffeomorphism constraint above depends on the angular partial derivatives of the radial shift, as pointed out also in  \cite{Bodendorfer:2015qie}; however, in that  analysis a further partial gauge fixing was introduced so that the radial shift does not depend on the angular coordinates and, therefore, no explicit form of the nonlocal terms was derived.

The Hamiltonian constraint splits into its Lorentzian and Euclidean parts. In the quantum theory, the Lorentzian part is traditionally treated by rewriting it in terms of commutators of the Euclidean part with the volume operator. Therefore let us focus here only on the Euclidean part $H_{\va E}$.
The extended Euclidean Hamiltonian can be written in the form
\be
\label{eq:ext:2}
\tilde{H}_{\va E}\approx H_{\va E}[N]+H_A[\gamma^A]+G_I[\gamma^I]\,,
\ee
where now
\ba
\label{eq:g:3}
\gamma^A&=&\int_r^{\bar{r}} dr'\Bigg[\frac{\partial_B}{E_3^r}\left(\frac{NE_I^{[A}E_J^{B]}\epsilon^{IJ}}{\sqrt{\text{det}(E)}}\right)\n\\
&+&\frac{E^A_I\partial_B}{(E_3^r)^2}\left(\frac{N\epsilon^{IJ}E_J^BE_3^r}{\sqrt{\text{det}(E)}}\right)-\frac{NE_I^A}{\sqrt{\text{det}(E)}}\left(\frac{E_J^B\delta^{IJ}A_B^3}{E_3^r}+A_r^I+\frac{\epsilon^{IJ}}{E_3^r}G_J\right)\n\\
&-&\frac{NE_I^AE_J^B\delta^{IJ}}{\sqrt{\text{det}(E)}(E_3^r)^2}\int_0^{r'}dr''\,H_B(r'')\Bigg]_{r'}\,,
\ea
and
\be
\label{eq:g:4}
\gamma^I=-\frac{\partial_A}{E_3^r}\left(\frac{N\delta^{IJ}E_J^AE_3^r}{\sqrt{\text{det}(E)}}\right)-\frac{N\epsilon^{IJ}E_J^AA_A^3}{\sqrt{\text{det}(E)}}\,.
\ee
Observe that, since the Hamiltonian is quadratic in the momenta, the second class constraints appear also as arguments of the smearings in \eqref{eq:g:3}.

As in the case of the radial diffeomorphisms, we can also define a {\it reduced Euclidean Hamiltonian} $\mathcal{H}_{\va E}$, neglecting the terms containing $A_A^3$ and $A_r^I$, explicitly
\ba
\label{eq:barh:2}
\mathcal{H}_{\va E}\approx\frac{\gamma}{\sqrt{\text{det}(E)}}\left(E_3^rE_I^A\epsilon\indices{^I_J}\partial_rA_A^J+E_I^AE_J^BA^I_{[A}A^J_{B]}+E_3^rE_I^AA_r^3A_A^I\right).
\ea
Then, the extended representation of the Euclidean Hamiltonian constraint, written in its unsmeared version, reads
\ba
\label{eq:hh:tilde}
\tilde{H}_{\va E}&\approx& \mathcal{H}_{\va E}\n\\
&+&\frac{\gamma}{\sqrt{\text{det}(E)}}\left\{E_I^AE_J^B\left[-\delta^{IJ}\frac{\mathcal{I}_A\mathcal{I}_B}{(E_3^r)^2}+\epsilon^{IJ}\partial_A\left(\frac{\mathcal{I}_B}{E_3^r}\right)\right]
-E_3^rE_I^A\left[\frac{\epsilon^{IJ}(\partial_BE^B_J)\mathcal{I}_A}{(E_3^r)^2}
+\partial_A\left(\frac{\epsilon^{IJ}E_J^B\mathcal{I}_B}{(E_3^r)^2}-\frac{\delta^{IJ}\partial_BE^B_J}{E_3^r}\right)\right]\right\}\,.\n\\
\ea
Notice that  $\text{det}(E)$ reduces to
\be
\label{eq:det}
\text{det}(E)\approx\frac{1}{2}\left(\epsilon_{rAB}\epsilon^{IJ}E^A_IE^B_J\right)E_3^r.
\ee

The expression \eqref{eq:hh:tilde} for the extended Euclidean Hamiltonian constraint, or equivalently \eqref{eq:ext:2}, \eqref{eq:g:3}, \eqref{eq:g:4}, represents the main result of this work.
%%%%%%%%%%%%

We could now try to compare our final result for the extended Hamiltonian constraint with the one obtained in \cite{Bodendorfer:2015aca} through the use of metric variables. However, as pointed out above, due to the imposition of the further gauge restriction $h_{rr}=1$ in \cite{Bodendorfer:2015aca}, we do not expect the final expressions to be equivalent. A possible way to see this is to restrict to the spherically symmetric case. If we replace the spherically symmetric connection and flux components that one can find for instance in  \cite{Bojowald:2005cb}, we can see that all the extra, nonlocal terms in \eqref{eq:hh:tilde} simplify, yielding the local term of the symmetry reduced  Hamiltonian encoding the connection component $A^3_\phi$; this is what we expected since this is the only connection component conjugate to a gauge fixed flux appearing in the spherically symmetric Hamiltonian constraint. On the other hand, due to the extra gauge condition $h_{rr}=1$, the spherically symmetric Hamiltonian derived in \cite{Bodendorfer:2015aca}  still contains nonlocal terms; only upon relaxing this constraint  one recovers the standard Hamiltonian of \cite{Kuchar:1994zk} and the two results match.

\section{Conclusions}\la{sec:Conc}

We have considered the canonical coordinates of GR phase space parametrazied by the Ashtekar-Barbero $SU(2)$ connection and its conjugate momentum  and introduced some partial gauge fixing conditions. These conditions bring in second class constraints in the theory, which we have implemented in our canonical analysis by means of the gauge unfixing procedure. As clarified above, such treatment of second class constraints is equivalent to the inversion of the Dirac matrix, as well as to the direct solution of the constraints; however, it presents the advantage that we can still use the standard Poisson bracket between the remaining (reduced) phase space coordinates, thus avoiding the complications of having to deal with the quantum representation of the Dirac bracket, at the price of introducing some nonlocal extra terms in the remaining first class constraints. We have thus shown that the Dirac program for second class constraint systems can be completed in order to reduce the gauge freedom of general relativity. 

This analysis lays the  basis  for the quantum description of black holes performed in \cite{Alesci:2018loi}. More precisely, the orthogonal gauge fixing performed here is useful to deal with 
the spherical symmetry reduction of a 3D spatial geometry. 
The strategy is to generalize techniques introduced for cosmological applications within the framework of Quantum Reduced Loop Gravity \cite{Alesci:2012md, Alesci:2013xd, Alesci:2014uha, Alesci:2014rra, Alesci:2015nja, Bilski:2015dra, Alesci:2016xqa, Alesci:2016gub} to impose the gauge fixing conditions in terms of expectation values on kinematical quantum states of the full theory. We can then use these reduced spin networks to build coherent states for a Schwarzschild quantum geometry, thus implementing the spherical symmetry reduction at the quantum level. The proper quantum dynamics  will be encoded in the operatorial version of the extended Euclidean Hamiltonian constraint \eqref{eq:hh:tilde} (and its Lorentzian contribution as well). Time evolution of the Schwarzschild geometry initial data according to resulting modified semiclassical Hamiltonian  is expected to generate an effective quantum corrected metric. 

Let us point out that, for the nice property of the volume operator to be diagonal with a simple spectrum on the quantum reduced states, as mentioned in  Sec.  \ref{sec:Intro} and at the base of all the great simplifications when dealing with the quantum constraint operators, it is crucial to employ the orthogonal gauge. This is a previous and separate step with respect to the symmetry reduction, which allows us to build reduced spin network basis states out of which coherent states can then be defined \cite{Alesci:2018loi}. In this sense, the classical analysis performed here for the GU procedure applied to the case of orthogonal gauge is a necessary step in order to then have a correct implementation of the remaining first class (extended) constraints, consistent with the quantum gauge reduction. This gives us access to technical tools crucial to go beyond the previous application of coherent state construction to the spherically symmetric case, see for instance  \cite{Dasgupta:2005yu}, where the difficulty to deal with the quantum dynamics (like, e.g.,  the explicit evaluation of the volume operator expectation value) prevented the derivation of an effective Hamiltonian coming from the full theory.

\addcontentsline{toc}{section}{Acknowledgement}
\section*{Acknowledgement}
We wish to thank Stefano Liberati for fruitful discussions at the initial stage of this project.
We acknowledge the John Templeton Foundation
for the supporting grant \#51876.
This work was supported in part by the NSF grant PHY-1505411, the
Eberly research funds of Penn State. This research was supported in part by Perimeter Institute for Theoretical Physics.
Research at Perimeter Institute for Theoretical Physics is supported in part by the Government of Canada through NSERC and by the Province of Ontario through MRI.
%%%%%%%%%%%%%%%%%%%%
\begin{appendix}
%\appendix
\section{Derivation of $\mathbb{A}^{-1}$}
\label{sec:app:a}
The matrix
\be
\mathbb{A}(\vec{x},\alpha)=
\begin{bmatrix}
c\indices{^A_J} & a\indices{^A_B}\\
b\indices{_I_J}& \emptyset\indices{_I_B}
\end{bmatrix}
\ee
has the symbolic structure
\be
A=
\begin{bmatrix}
c & a\\
b & 0
\end{bmatrix}\,.
\ee

If $a,b,c$ were just numbers, the inverse would be
\be
\label{eq:amo}
A^{-1}=
\begin{bmatrix}
0 & b^{-1}\\
a^{-1} & -a^{-1}c\,b^{-1}
\end{bmatrix}\,.
\ee
We must then find a distributional equivalent of \eqref{eq:amo}. The equivalent of $a^{-1}$ is a distribution $(a^{-1})\indices{^A_B}(\vec{x},\vec{y})$ such that
\be
\int d\vec{y}\,(a^{-1})\indices{^A_B}(\vec{x},\vec{y})\,a\indices{^B_C}(\vec{y})=\delta^A_C\,\alpha(\vec{x})\,.
\ee
Similarly,
\be\la{eq:bm0}
\int d\vec{y}\,(b^{-1})\indices{^I^J}(\vec{x},\vec{y})\,b\indices{_J_K}(\vec{y})=\delta^I_K\,\alpha(\vec{x})\,.
\ee
From \eqref{eq:small:a}-\eqref{eq:small:b}, and from the fact that $\alpha(\vec{x})$ is a smearing function obeying vanishing boundary conditions, it is straightforward to verify that the expressions \eqref{eq:small:amo}-\eqref{eq:small:bmo} are the correct inverses.

The distributional equivalent of $ -a^{-1}c\,b^{-1}$ is the matrix
\be
\label{eq:dmo2}
\begin{split}
&d\indices{^A^J}(\vec{x},\vec{y})=\\
&=-\int d\vec{z} \int d\vec{w}\,(a^{-1})\indices{^A_B}(\vec{x},\vec{w})\,\tilde{c}\indices{^B_I}(\vec{w},\vec{z})(b^{-1})\indices{^I^J}(\vec{z},\vec{y})
\end{split}
\ee
where $\tilde{c}\indices{^A_I}$ is the distributional matrix such that
\be
\int d\vec{y}\,\tilde{c}\indices{^A_I}(\vec{x},\vec{y})\,\alpha(\vec{y})=c\indices{^A_I}(\vec{x})\,.
\ee
Then, from \eqref{eq:small:c},
\be
\label{eq:small:c:2}
\tilde{c}\indices{^A_I}(\vec{x},\vec{y})=\epsilon\indices{_I^J}E_J^A(\vec{x})\,\delta(\vec{x},\vec{y})\,.
\ee
Inserting \eqref{eq:small:c:2} into \eqref{eq:dmo2}, and using \eqref{eq:amo}-\eqref{eq:bm0}, it is immediate to show that $d\indices{^A^J}(\vec{x},\vec{y})$ corresponds to Eq.\,\eqref{eq:dmo}.

This completes our derivation of $\mathbb{A}^{-1}$.
\end{appendix}
%%%%%%%%%%%%%%%%%%%%%%%%%%%%%%%%%%

%\bibliography{spherical}


\end{document}